\documentclass[letterpaper,12pt]{article}
\usepackage{amsmath}
\usepackage{amsthm}
\usepackage{amssymb}
\usepackage{amsfonts}
\usepackage{dsfont}
\usepackage{mathptmx}
\usepackage{setspace}
\usepackage{graphics}
\usepackage{amsmath}
\usepackage[authoryear,comma,sectionbib]{natbib}

\usepackage[dvips]{graphicx}
\usepackage[colorlinks=true,  linkcolor=blue, urlcolor=blue, citecolor=blue]{hyperref}
\usepackage{booktabs} % for the bold lines above and below the fancy tables
\begin{document}

\title{Adjusted Plus-Minus for NHL Players using Ridge Regression with Goals, Shots, Fenwick, and Corsi}
\author{\large Brian Macdonald\footnote{email: \url{bmac@jhu.edu}}
}
\date{{\footnotesize\today}}

\maketitle

\begin{abstract}
Regression-based adjusted plus-minus statistics were developed in basketball and have recently come to hockey. The purpose of these statistics is to provide an estimate of each player's contribution to his team, independent of the strength of his teammates, the strength of his opponents, and other variables that are out of his control.  One of the main downsides of the ordinary least squares regression models is that the estimates have large error bounds.  Since certain pairs of teammates play together frequently, collinearity is present in the data and is one reason for the large errors. In hockey, the relative lack of scoring compared to basketball is another reason. To deal with these issues, we use ridge regression, a method that is commonly used in lieu of ordinary least squares regression when collinearity is present in the data.  We also create models that use not only goals, but also shots, Fenwick rating (shots plus missed shots), and Corsi rating (shots, missed shots, and blocked shots).  One benefit of using these statistics is that there are roughly ten times as many shots as goals, so there is much more data when using these statistics and the resulting estimates have smaller error bounds. The results of our ridge regression models are estimates of the offensive and defensive contributions of forwards and defensemen during even strength, power play, and short handed situations, in terms of goals per 60 minutes. The estimates are independent of strength of teammates, strength of opponents, and the zone in which a player's shift begins. 
\end{abstract}


\tableofcontents
\listoftables
\listoffigures

\section{Introduction}
    Though the plus-minus statistic was first used in hockey, advanced versions of plus-minus have been developing more quickly in basketball.  These new versions attempt to correct one or more of the problems associated with the traditional plus-minus statistic, which depends heavily on the strength of a player's teammates and opponents, and on other variables out of a player's control.  Regression-based versions of adjusted plus-minus ($APM$) statistics for NBA players can be found in  \cite{winston}, \cite{rosenbaum}, \cite{lewin}, \cite{eli}, \cite{ilardibarzilai}, \cite{sill}, and \cite{fearnhead-taylor-nba}.
    
    In \cite{apm} and \cite{apm2}, the author developed similar models for hockey.  In \cite{apm}, the author used weighted least squares models similar to those in \cite{rosenbaum} and \cite{ilardibarzilai} to find the estimates of each player's offensive and defensive contribution during even strength situations, adjusted for the strength of his teammates and opponents.  The contributions are given in terms of goals per 60 minutes and goals per season.  Special teams situations are addressed in \cite{apm2}.  Information about the zone in which each shift begins was also used in \cite{apm2} in order to get estimates that are independent of the zone on the ice in which a player typically begins his shifts.    
    
    In many of the basketball articles, and also in the hockey articles \cite{apm} and \cite{apm2}, it was noted that one downside of the ordinary least squares regression models is that the results can have large error bounds, which are measures of uncertainty in the estimates.  Since two main uses of these estimates could be (1) deciding which players to trade for and (2) establishing parameters for salary negotiations, smaller errors, and hence more precise estimates, have significant value to NHL analysts and decision-makers.

    One reason for the large errors is the high collinearity in the data caused by teammates who play together frequently, a common occurrence in many sports.  For example, Henrik and Daniel Sedin, twin brothers who play for the Vancouver Canucks, are almost always on the ice together.  A regression model will have a difficult time telling them apart (both statistically and biologically) and their estimates tend to have large errors.  In an extreme case where two players always play together, the ordinary least squares estimates will not even be unique.
    
    Another reason for the high errors in hockey is the relative lack of scoring when compared to a sport like basketball.  A typical hockey team only scores two to four goals per game on average during a season. The low goal scoring rates, coupled with randomness and luck involved with goal scoring, makes it difficult to properly judge players using goals alone without using multiple seasons of data.  Additionally, a player's goals for and goals against while he is on the ice is dependent on the quality of goalies on the ice.  Ideally, one would prefer to estimate a player's abilities in a way that is independent of the quality of the goalies he faces, and independent of the quality of goalies on his team.
    
    \subsection{Brief summary of the new models}
    In light of these observations, we make two modifications to the models given in \cite{apm} and \cite{apm2}.  First, in lieu of ordinary least squares regression models, we use ridge regression models (\cite{hoerl-ridge}, \cite{hoerl-kennard-ridge}), similar to the model used for basketball in \cite{sill}.  Ridge regression frequently reduces the error bounds in the estimates and improves the predictive performance of the model when collinearly exists in the data.  Ridge regression introduces bias in the estimates, but the tradeoff is typically worthwhile.  The model is discussed in detail in Section \ref{ridge}.

    The second change we make is to form additional models that use three 
    other statistics, in addition to goals, as the dependent variable.  These additional models use shots, Fenwick rating (shots plus missed shots), and Corsi rating (shots, missed shots, and blocked shots).  These statistics were chosen because each of them has been shown to be very good indicators of performance at the team level (\cite{shots-fwick-corsi}, \cite{possessioniseverything}).  
    
    There are pros and cons to using these statistics, and that is one reason that we will use them \textit{in addition to} goals and not \textit{instead of} goals.  For example, on one hand, shots, Fenwick rating, and Corsi rating ignore the shooting ability of players, although many hockey analysts would argue that a player's shooting ability is not nearly as significant as his ability to generate shots.  Also, some would argue that missed shots and blocked shots should not be included or should not be considered good, since they are attempted shots that never reached the goal.  However, if a team has more shots, missed shots, and blocked shots than their opponents, it is most likely an indication of a territorial advantage and an advantage in terms of puck possession.  In order to take a shot, a player must possess the puck, and typically that player is also in the offensive zone.  
    
    The relationships among goals, shots, Fenwick rating, and Corsi rating are described well in \cite{shots-fwick-corsi} and discussed further in \cite{spm}.  In both articles, the authors show that shots, Fenwick rating, and Corsi rating are better indicators than goals of a team's future performance when one uses data from only half of a season.  Based on this analysis, we believe the results based on shots, Fenwick rating and Corsi rating do have value, especially for our models that are based on only one season's worth of data.  The reader can decide for him- or herself how much value those results have.
    
    One nice benefit of using of these additional statistics is that they are far more prevalent than goals.  Typically, there are roughly 10 shots to every goal.  The extra data goes a long way to producing estimates with much smaller error bounds.  Also, for the most part, those statistics are independent of goalies, so the strength of the goalies on a player's team will not have much of an affect on the estimates of his contributions.  When using goals, the estimate of a player's defensive contribution, in particular, can be positively or negatively affected by the performance of the goalie playing behind him.  
     
     In order to more easily compare the results based on these additional statistics with the results based on goals, the new results were rescaled using league average shooting percentages.  By shooting percentages we mean goals per shot $\left(\frac{Goals}{Shots}\right)$, goals per Fenwick rating $\left(\frac{Goals}{Shots + Missed \, Shots}\right)$, and goals per Corsi rating $\left(\frac{Goals}{Shots+ Missed \, Shots + \, Blocked \, Shots}\right)$.  League averages of these shooting percentages were computed for even strength, power play, and short handed situations separately, using data from the last four full NHL seasons.  
     
     The results based on shots, Fenwick rating, and Corsi rating were then rescaled by multiplying by the league average goals per shot, goals per Fenwick rating, and goals per Corsi rating, respectively.  These results are in the units of expected goals per 60 minutes based on shots, Fenwick, or Corsi.  A player's rescaled results based on shots can be thought of as his contribution to his team, in the units of expected goals per 60 minutes, based on shots for and shots against when he was on the ice.  The results remain independent of the strength of his teammates, the strength of his opponents, and the zone in which his shifts begin.  The rescaled results based on Fenwick rating and Corsi rating can be interpreted  similarly.

    We use four separate ridge regression models for even strength situations using each of these four statistics (goals, shots, Fenwick rating, and Corsi rating) as the response variable.  Each even strength model gives an even strength offensive and defensive component of $APM$ for each player in terms of goals per 60 minutes or expected goals per 60 minutes.  These components can be added to give a player's total contribution at even strength in terms of goals per 60 minutes.   We also have four separate models for special teams situations, one for each of the four statistics.  Each special teams model gives an offensive and defensive component on the power play, as well as an offensive and defensive component during short handed situations, in terms of goals per 60 minutes.  In total, we get 36 estimates for each player in terms of goals per 60 minutes.  If the results are expressed in terms of goals per season, then even strength, power play, and shorthanded results can be added to give estimates of offensive, defensive, and total contributions in all situations, in terms of goals per season.  So, in this case, we get 48 different ratings for each player.  This can be a bit of information overload, and when we present the results here, we will need to be selective regarding which components of $APM$ are listed.  Notation will be important as well.
    
    \subsection{Notation}
    Notation for the offensive, defensive, and total contribution of a player (forward or defensemen) during even strength, power play, and short handed situations, using the model with goals as the response variable, is given in Table \ref{notation}.  
        \begin{table}[h]
                \centering
                    \caption[Summary of notation for $APM$ results using goals.]{Summary of notation for $APM$ results using goals. For each player (forward or defensemen), we have offensive, defensive, and total contributions during even strength, power play, and short handed situations, in terms of goals per season.}
                    \begin{tabular}{lrrr}
                    \\
                        \toprule
                        Strength &  Offense    & Defense    &    \quad Total \\
                        \midrule
                        Even strength    & $G^\text{off}_{EV}$ & $G^\text{def}_{EV}$ & $G_{EV}$ \\
                        \addlinespace[.5em]     
                        Power play       & $G^\text{off}_{PP}$  & $G^\text{def}_{PP}$ & $G_{PP}$ \\
                        \addlinespace[.5em]     
                        Short handed     & $G^\text{off}_{SH}$  & $G^\text{def}_{SH}$ & $G_{SH}$ \\
                        \addlinespace[.5em]     
                        All situations   & $G^\text{off}_{}$  & $G^\text{def}_{}$ & $G_{\hskip .46cm}$ \\
                        \bottomrule
                    \end{tabular}
                    \label{notation}
                \end{table}
    The adjusted plus-minus results based on shots, Fenwick rating and Corsi rating are denoted similarly, except with  ``\textit{S}'', ``\textit{F}'', and ``\textit{C}'', respectively, instead of the ``\textit{G}'' that is used for goals. 
    For example, the \textbf{ev}en strength \textbf{off}ensive component of $APM$ using \textbf{g}oals, \textbf{s}hots, \textbf{F}enwick rating, and \textbf{C}orsi rating are denoted $G^\text{off}_{EV}, S^\text{off}_{EV}, F^\text{off}_{EV}$ and $C^\text{off}_{EV}$, respectively.  The per 60 minute versions of these statistics are denoted similarly, but with a subscript of ``$60$'' included.  For example, \textbf{ev}en strength \textbf{off}ensive component of adjusted plus-minus per \textbf{60} minutes using \textbf{g}oals is denoted $G^\text{off}_{EV,60}$.  

    \subsection{Example of the results} 
    In Table \ref{offense}, we give an example of the results.  We list the top 10 players in offense during the 2007-08, 2008-09, 2009-10, and 2010-11 seasons according to $G^\text{off}$, the offensive component of $APM$ in terms of goals per season.  We also list the players' offensive contributions according to the $APM$ models based on the other statistics.  

% latex table generated in R 2.13.1 by xtable 1.6-0 package
% Sat May 05 17:32:42 2012
\begin{table}[h!]
\begin{center}
\caption{The top $10$ offensive players in the NHL according to $G^{\text{off}}$.}
\label{offense}
{\footnotesize
\begin{tabular}{rlllrrrrrrrr}
  \addlinespace[.3em] \toprule 
 & Player & Pos & Team & $G^{\text{off}}$ & $S^{\text{off}}$ & $F^{\text{off}}$ & $C^{\text{off}}$ & $G^{\text{off}}_{EV,60}$ & Err & $S^{\text{off}}_{EV,60}$ & Err \\ 
  \midrule 
1 & Sidney Crosby & C & PIT & 23 & 12 & 13 & 14 & 0.83 & 0.20 & 0.42 & 0.07 \\ 
  2 & Jonathan Toews & C & CHI & 18 & 8 & 8 & 9 & 0.45 & 0.20 & 0.22 & 0.07 \\ 
  3 & Alex Ovechkin & LW & WSH & 17 & 17 & 20 & 24 & 0.46 & 0.18 & 0.45 & 0.07 \\ 
  4 & Daniel Sedin & LW & VAN & 16 & 13 & 13 & 15 & 0.47 & 0.18 & 0.44 & 0.08 \\ 
  5 & Joe Thornton & C & S.J & 16 & 11 & 11 & 15 & 0.34 & 0.18 & 0.26 & 0.06 \\ 
  6 & Nicklas Backstrom & C & WSH & 16 & 11 & 12 & 14 & 0.23 & 0.19 & 0.28 & 0.07 \\ 
  7 & Evgeni Malkin & C & PIT & 15 & 11 & 11 & 12 & 0.40 & 0.20 & 0.31 & 0.06 \\ 
  8 & Ryan Getzlaf & C & ANA & 15 & 6 & 8 & 9 & 0.31 & 0.19 & 0.07 & 0.07 \\ 
  9 & Pavel Datsyuk & C & DET & 15 & 10 & 11 & 12 & 0.53 & 0.19 & 0.27 & 0.07 \\ 
  10 & Jason Spezza & C & OTT & 13 & 7 & 8 & 9 & 0.37 & 0.21 & 0.25 & 0.07 \\ 
   \bottomrule 
\end{tabular}
}
\end{center}
\end{table}
     Recall that $S^\text{off}$, $F^\text{off}$, and $C^\text{off}$ have been rescaled by multiplying by the league average goals per shot, goals per Fenwick rating, and goals per Corsi rating, respectively.  Note that these statistics are in the units of goals per season or expected goals per season based on shots, Fenwick, or Corsi, so they do depend on playing time.  Sidney Crosby, for example, has missed significant time in two of the four seasons, and that has a big impact on his rating, although he still leads the league in $G^\text{off}$ by a sizeable margin.  
     
     Some per 60 minute results, $G^\text{off}_{EV,60}$ and $S^\text{off}_{EV,60}$, along with their standard errors, are given in the last four columns of that table.  These statistics are independent of playing time, so they do not depend on how much their coaches play them.  They also do not depend on how much time these players have spent on injured reserve.  We believe that both the per season and per 60 minutes versions of these statistics have value, and we will continue to list both versions in our tables.  
    
    In this paper, we will mostly give results based on models that contain data from four NHL seasons: 2007-08, 2008-09, 2009-10, and 2010-11.  However, since we are now using ridge regression, single season results are stable enough to have value.  One might prefer to see a player's progression from season to season rather than seeing a single number for all four years.  Also, one might prefer to make adjustments so that the statistics for a player are relative to a replacement player at the same position.  An example of Sidney Crosby's $APM$ statistics in each of the past four seasons, with adjustments for position and replacement players, is given in Table \ref{yearlyresults}.  We have also included his 4-year results for comparison.
        \begin{table}[h]
            \centering
            \caption[Sidney Crosby's $APM$ statistics during 2007-11 using goals.]{Sidney Crosby's $APM$ statistics over the past four seasons using goals.}
            \begin{tabular}{@{\extracolsep{-.25ex}}lrrrrrrrrrrrr}
            \toprule
            Year& Age& 	GP&	\quad $G^\text{off}$ &G$^\text{def}$& $G$&\quad $G^\text{off}_{EV}$ & $G^\text{def}_{EV}$ & $G^{}_{EV}$ & \quad $G^\text{off}_{PP}$ & $G^\text{def}_{PP}$ & $G^{}_{PP}$\\
            \midrule
            2007& 20&	53&	\quad 25&	9&	33&	\quad 19&	7&	26&	\quad 6&	2&	7\\
            2008& 21&	77&	\quad 30&	3&	33&	\quad 21&	2&	23&	\quad 9&	1&	10\\
            2009& 22&   81&	\quad 37&	4&	41&	\quad 31&	2&	34&	\quad 5&	2&	7\\
            2010& 23&	41&	\quad 17&	1&	18&	\quad 13&	0&	13&	\quad 3&	1&	5\\
            4-yr& 20-23&	\quad 63&	29&	1&	\quad 30&	23&	1&	\quad 24&	5&	1&	6\\
            \bottomrule
            \end{tabular}\label{yearlyresults}
        \end{table}
        
    We also note that the errors in our estimates are lower than those reported in \cite{apm} and \cite{apm2}, where the author used ordinary least squares (OLS) regression instead of ridge regression.  As an example, we give Alex Ovechkin's even strength offensive contributions per 60 minutes in Table \ref{errors}, along with their standard errors.
     The errors in Ovechkin's $G^\text{off}_{EV,60}$ are smaller than those reported in \cite{apm} and \cite{apm2}.  Also, the errors in Ovechkin's $S^\text{off}_{EV,60}$, $F^\text{off}_{EV,60}$, and $C^\text{off}_{EV,60}$ are smaller than the errors in $G^\text{off}_{EV,60}$.  
          This trend can also be seen in Table \ref{offense}.  The standard errors in $S^\text{off}_{EV,60}$ are much lower than the standard errors in $G^\text{off}_{EV,60}$ for all of the players in that table.  The errors are still not small enough to be ignored, as the confidence intervals of many of the estimates still overlap.  Nevertheless, the $APM$ estimates with smaller error bounds, coupled with the additional $APM$ estimates based on shots, Fenwick rating, and Corsi rating, are useful metrics with which to analyze the performance of NHL players. 
    % latex table generated in R 2.13.1 by xtable 1.6-0 package
         % Sat Dec 31 10:53:58 2011
         \begin{table}[h!]
         \begin{center}
         \caption{Alex Ovechkin's EV offense statistics, with standard errors.}
         \label{errors}
         {\footnotesize
         \begin{tabular}{rlrrrrrrrrr}
           \addlinespace[.3em] \toprule 
         Player & Pos & Team & $G^\text{off}_{EV,60}$ & $Err$ & \quad $S^\text{off}_{EV,60}$ & $Err$ & \quad  $F^\text{off}_{EV,60}$ & $Err$ & \quad $C^\text{off}_{EV,60}$ & $Err$ \\  \midrule 
         Alex Ovechkin & LW & WSH & 0.46 & 0.18 & \quad 0.45 & 0.07 & \quad 0.53 & 0.06 & \quad 0.63 & 0.05 \\ 
            \bottomrule 
         \end{tabular}
         }
         \end{center}
         \end{table}
         
    The rest of this paper is organized as follows.  First, we describe the ridge regression models in detail in Section \ref{ridge}.  In Section \ref{results}, we give the players that $APM$ determines as the Hart Trophy, Norris Trophy, and Selke Trophy finalists (most valuable player, best defensemen, and best defensive forward, respectively) during the 2007-08, 2008-09, 2009-10, and 2010-11 seasons combined.  We finish with conclusions and future work in Section \ref{conclusion}.  In the Appendix, we give a brief comparison of ordinary least squares and ridge regression, and describe how we chose our ridge parameter in our ridge regression models.

\section{Ridge Regression Model}\label{ridge}
\begin{figure}[h!]
            \centering
            \includegraphics[width=.45\textwidth]{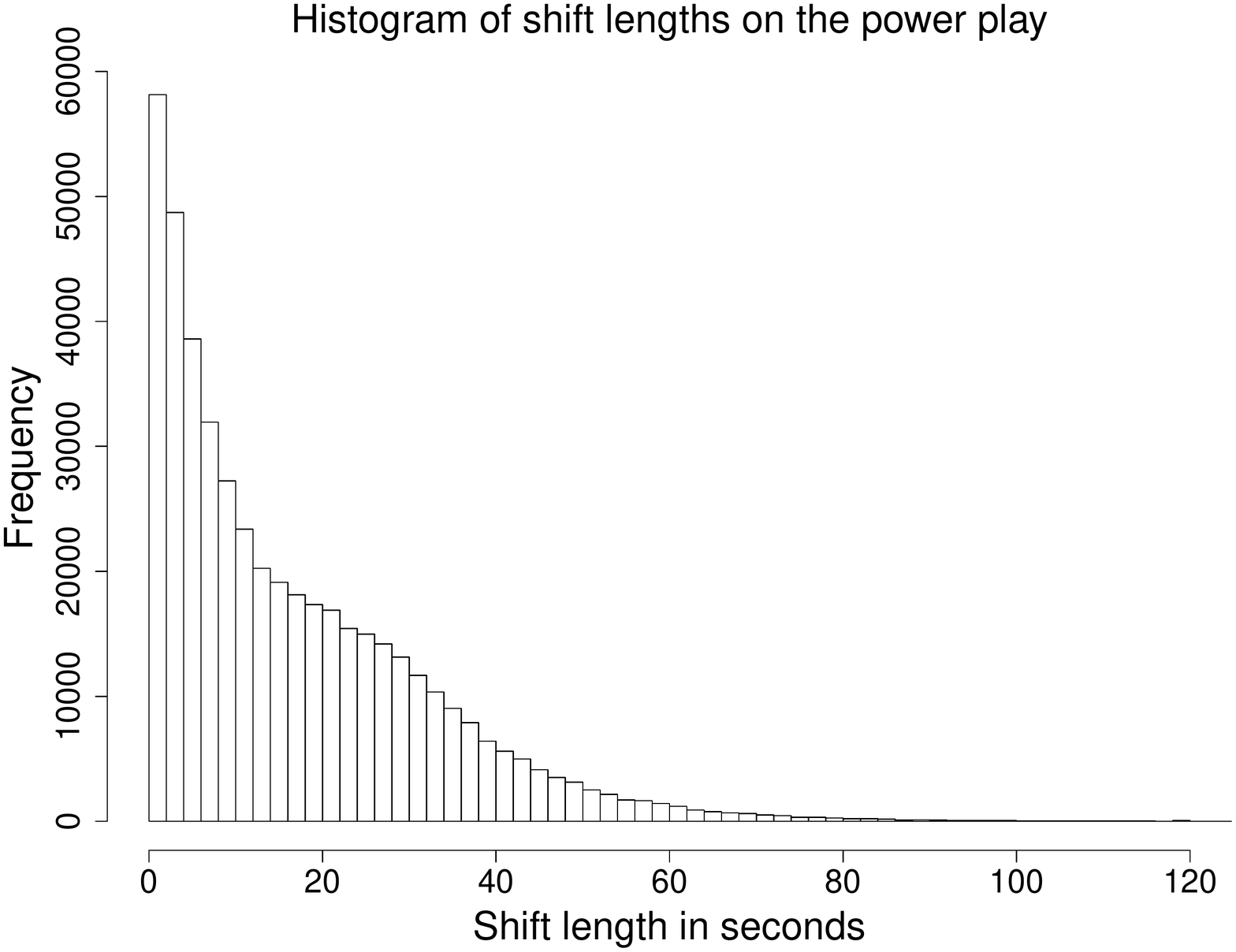}
            \includegraphics[width=.45\textwidth]{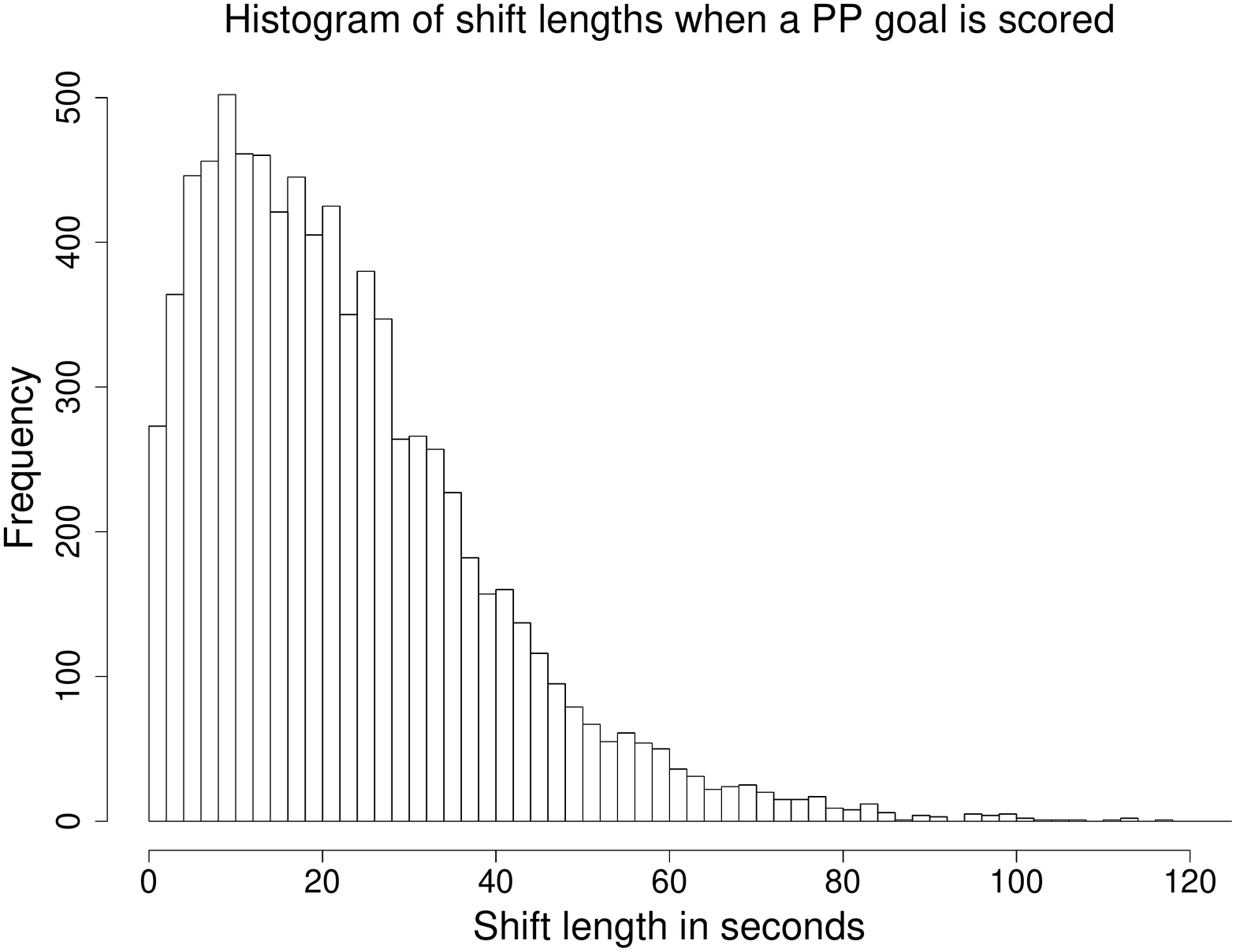}
            \caption[Shift lengths for all PP shifts and PP shifts when a goal is scored]{A comparison of the shift lengths during power play situations for all shifts (left) and only shifts during which a goal is scored (right).  Typically, shift lengths are longer for the shifts when a goal is scored.  This observation is similar to that made by \cite[Figure 6]{thomas-ventura} for even strength situations.}
            \label{shiftlengths}
            \end{figure}

\begin{figure}[h!]
            \centering
            \includegraphics[width=.45\textwidth]{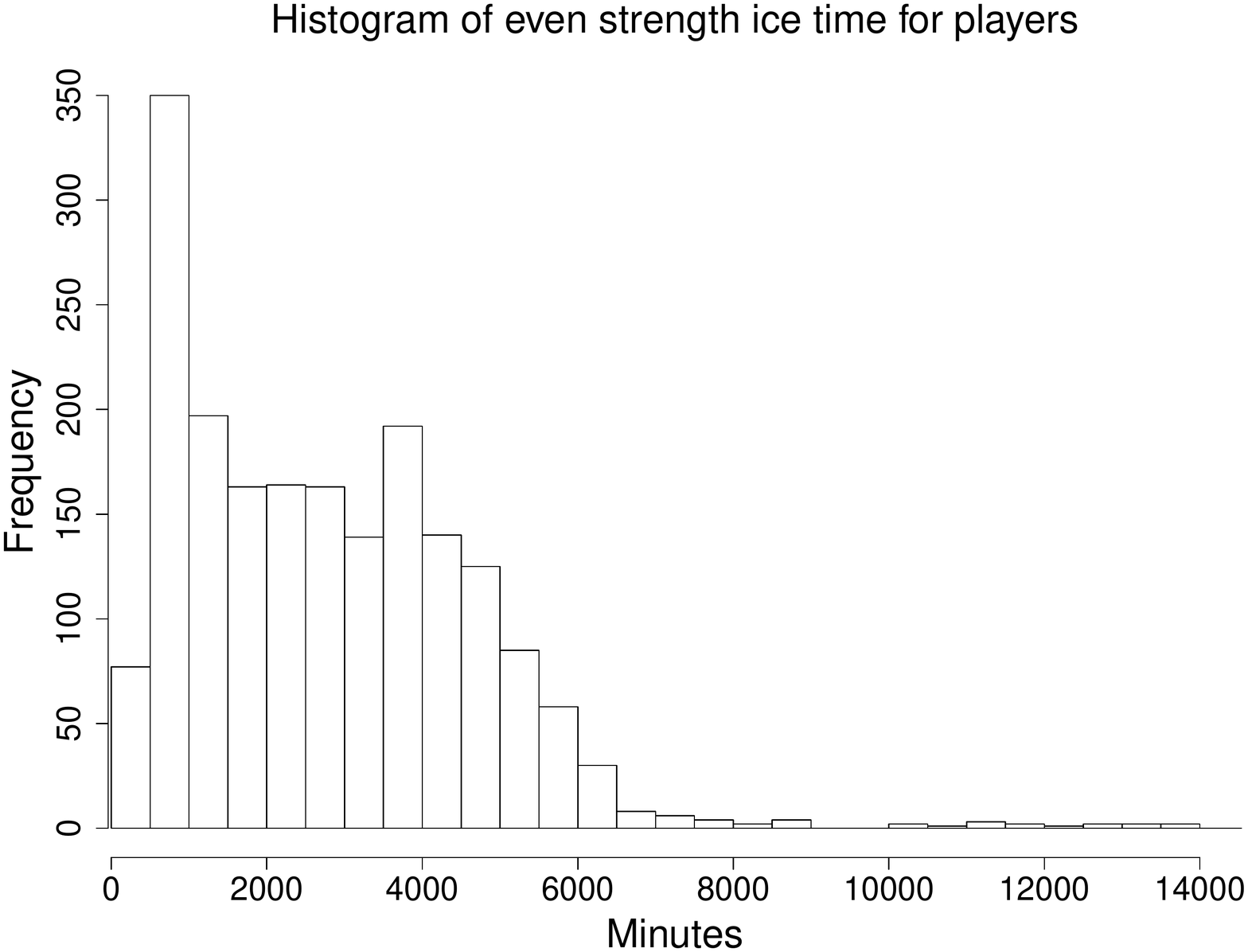}
            \includegraphics[width=.45\textwidth]{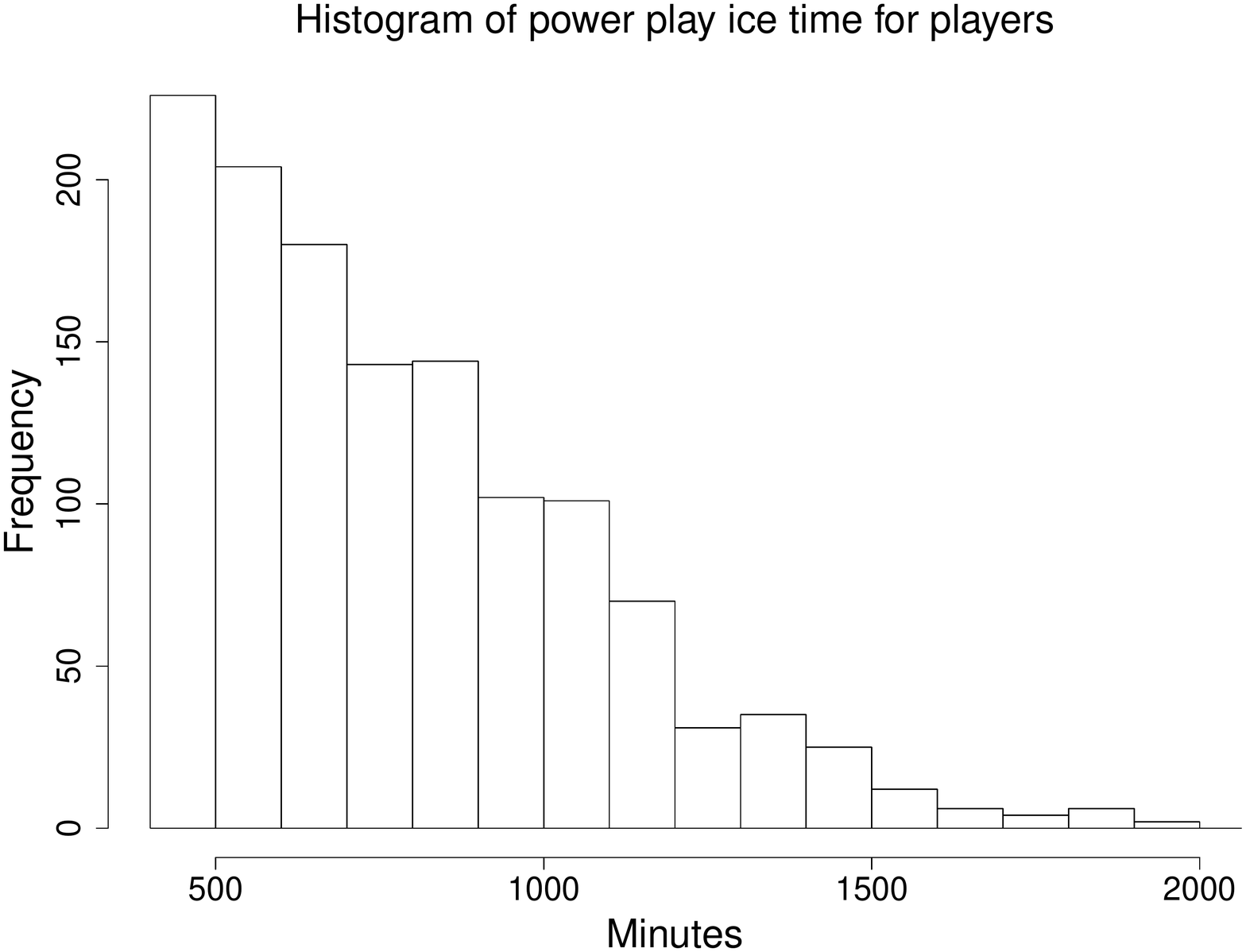}
            \caption[Distribution of players' ice time for EV and PP situations]{Distribution of players' ice time during even strength (left) and power play (right) situations.  The small grouping of players with more than 10,000 minutes of even strength playing time are all goalies.}
            \label{icetime}
            \end{figure}

    We now describe the setup of our model.  We use information about the players on the ice during every shift of every game during the 2007-08, 2008-09, 2009-10, and 2010-11 seasons, as well as the outcome of each shift. We divide this data into even strength and special teams situations, and we remove empty net situations from both data sets.  Each shift gives two lines of data: one line corresponding to the goals per 60 minutes scored by the home team, and one line corresponding to the goals per 60 minutes scored by the away team.  Both of these observations are weighted by the duration of that shift. We denote the total number of observations by $N$.  For even strength, we have $N=2,324,528,$ while for special teams situations, we have $N = 461,022$.  We note that the average duration of a shift is 4.5 seconds longer for special teams than for even strength.  Other observations about shift lengths and ice time for players, along with accompanying figures, can be found in Figures \ref{shiftlengths} and \ref{icetime} the Appendix.
    
    Let $J$ denote the number of players in the league, let $y$ denote the goals (or shots, Fenwick rating, or Corsi rating) per 60 minutes during an observation, and let $X_j$ and $D_j$ be indicator variables that are defined as follows:
        \begin{align}\label{ibvariables} \notag
            X_j &= \left\{
                    \begin{array}{ll}
                      1, & \hbox{ skater $j$ is on offense during the observation;} \\
                      0, & \hbox{ skater $j$ is not playing or is on defense during the observation;}\notag
                    \end{array}
                  \right.
            \\
            D_j &= \left\{
                    \begin{array}{ll}
                      1, & \hbox{ skater $j$ is on defense during the observation;} \\
                      0, & \hbox{ skater $j$ is not playing or is on offense during the observation;}
                    \end{array}
                  \right.
        \end{align}
        where $1 \leq j \leq J.$  Note that by ``skater" we mean a forward or a defensemen, but not a goalie.   We also note that for the models which use goals, we included defensive variables for goalies.  Let $Z_{off}$ and $Z_{def}$ be indicator variables defined as follows:
            \begin{align}\label{zonevariables} \notag
                    Z_{off} &= \left\{
                            \begin{array}{ll}
                              1, &  \hbox{ the observation corresponds to a shift that begins with a faceoff in } \\
                                  &  \hbox{ the offensive zone,} \\
                              0, & \hbox{ otherwise}
                            \end{array}
                          \right.
                          \\
                    Z_{def} &= \left\{
                            \begin{array}{ll}
                              1, & \hbox{ the observation corresponds to a shift that begins with a faceoff in } \\ 
                                  & \hbox{ the defensive zone,} \\
                              0, & \hbox{ otherwise}
                            \end{array}
                          \right.        
            \end{align}

        To clarify, we give an example.  Suppose that in one shift, skaters 1-5 are on the ice for the home team, and skaters 6-10 are on the ice for the away team.  Suppose that this is a shift of duration $t_1$ seconds, and that the home team scores a goal during this shift.  For this shift we would have two lines of data, one for goals per 60 minutes scored by the home team, and the second for goals per 60 minutes scored by the away team.  These two rows of data would look like this:
                \begin{align*}
                     X &= [1\,\,1\,\,1\,\,1\,\,1\,\,0\,\,0\,\,0\,\,0\,\,0\,\,0\,\cdots 0],
                     D  = [0\,\,0\,\,0\,\,0\,\,0\,\,1\,\,1\,\,1\,\,1\,\,1\,\,0\,\cdots 0],  \quad y = \frac{1}{t_1}\cdot 3600, 
                               \\
                     X &= [0\,\,0\,\,0\,\,0\,\,0\,\,1\,\,1\,\,1\,\,1\,\,1\,\,0\,\cdots 0],
                     D  = [1\,\,1\,\,1\,\,1\,\,1\,\,0\,\,0\,\,0\,\,0\,\,0\,\,0\,\cdots 0],  \quad y = \frac{0}{t_1} \cdot 3600.
                \end{align*} 
        We note that $\frac{1}{t_1}$ is in the units of goals per second, so we multiple by 3,600 to get goals per 60 minutes.  
        For even strength situations, we start with the following linear model:
                    \begin{align}\label{model1eq}
                        y = \beta_0 + \beta_1 X_1 + \cdots + \beta_J X_J &+ \delta_1 D_1 + \cdots + \delta_J D_J + \zeta_{off} Z_{off} + \zeta_{def} Z_{def}.
                    \end{align}
        The quantities of interest are the $\beta_j$s and $\delta_j$s, which are player $j$'s offensive and defensive contributions, respectively, in terms of goals per 60 minutes.  The coefficients $\zeta_{off}$ and $\zeta_{def}$ can be regarded as estimates of the value of starting a shift in the offensive or defensive zone, respectively, in terms of goals per 60 minutes.  For special teams situations, we start with a model that is similar to \eqref{model1eq} and is described in \cite{apm2}.  In total, there are 8 models: an even strength model and a special teams model for each of the four statistics. 
        
        A linear model like \eqref{model1eq} can also be expressed as a system of linear equations in matrix form as
            \begin{align}\label{matrixform}
                y = X \beta,
            \end{align}
        where $y$ is an $N\times 1$ vector of response variables, $X$ is an $N\times (2J+3)$ matrix of the explanatory variables, and $\beta$ is an $(2J+3) \times 1$ vector of coefficients, the quantities we are interested in.  Typically, when the number of observations, $N$, is much greater than the number of explanatory variables, $2J+3$, no solution to \eqref{matrixform} exists, and one must find some sort of ``best fit'' solution.  
        
        Instead of using OLS as in \cite{apm} and \cite{apm2} to find the best fit, we use ridge regression.  For the sake of those readers who are unfamiliar with ridge regression, we  give a brief description of how to find the best fit estimates using OLS regression and ridge regression, and how the two methods are related, in the Appendix.  We also discuss how we chose the ridge parameter $\lambda$ in that section.
        
        \subsection{Differences in OLS and Ridge Estimates} 
\begin{figure}[h]
            \centering
            \includegraphics[width=.75\textwidth]{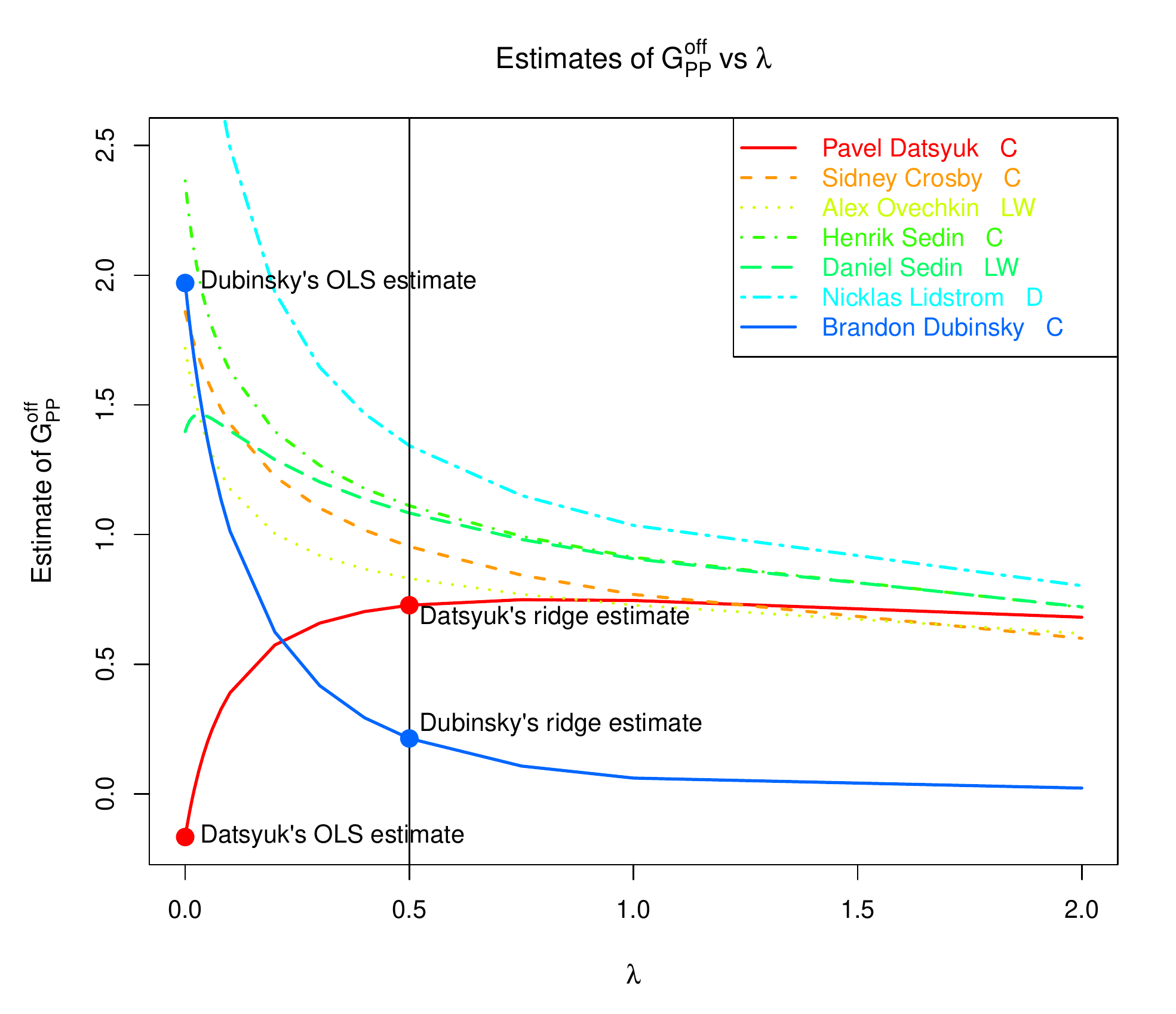}
            \caption{Estimates of $G^\text{off}_{PP,60}$ for some players, for different values of $\lambda$.}
            \label{tracecurves}
            \end{figure}

        The effect that this ridge parameter $\lambda$ has on the estimates can be seen in Figure \ref{tracecurves}.  
        In this example, we plot the estimated coefficient for $G^\text{off}_{PP,60}$ (offensive contribution on the power play in terms of goals per 60 minutes) of a few players in the league for different choices of $\lambda.$  Note that when $\lambda=0$, Pavel Datsyuk (solid red line) actually has a negative estimate, and Brandon Dubinsky (solid blue line) has a very high positive estimate in line with the league's elite offensive players.  Dubinsky is a valuable offensive player, but one would not expect his rating to be that much higher than Datsyuk's rating or among the league's elite.  Also, we would not consider Datsyuk to be a below average player on the power play.  We note that $\lambda=0$ corresponds to the ordinary least squares estimates, so these are the estimates we would have gotten for Dubinsky and Datsyuk if we had not used ridge regression.  
                
        However, notice that for larger choices of $\lambda$, the estimates begin to stabilize.  Datsyuk's estimate moves towards the estimates of the league's elite players, while Dubinsky's estimate returns to a more reasonable level.  These estimates agree with most people's intuition that Datysuk is an elite offensive player, while Dubinsky is an above average offensive player, but not an elite player as his ordinary least squares estimate suggested.  
        
        For Dubinsky, the unexpected result for $\lambda=0$ is probably due to minimal playing time.  For Datsyuk, it is probably due to the fact that he spent 90\% of his power play time with one of his teammates, Nicklas Lidstrom.  While Datsyuk's estimate starts below zero for $\lambda=0$ and increases as $\lambda$ increases, Lidstrom's estimate (dotted and dashed, light blue line) is off the figure near 4.0 for $\lambda=0$, and rapidly decreases as $\lambda$ increases.  While we would expect Lidstrom to have a good offensive rating on the power play, 4.0 is unusually high, and the ridge regression seems to be tempering Lidstrom's estimate while correcting Datsyuk's.  
        
\begin{figure}[h]
            \centering
            \includegraphics[width=.75\textwidth]{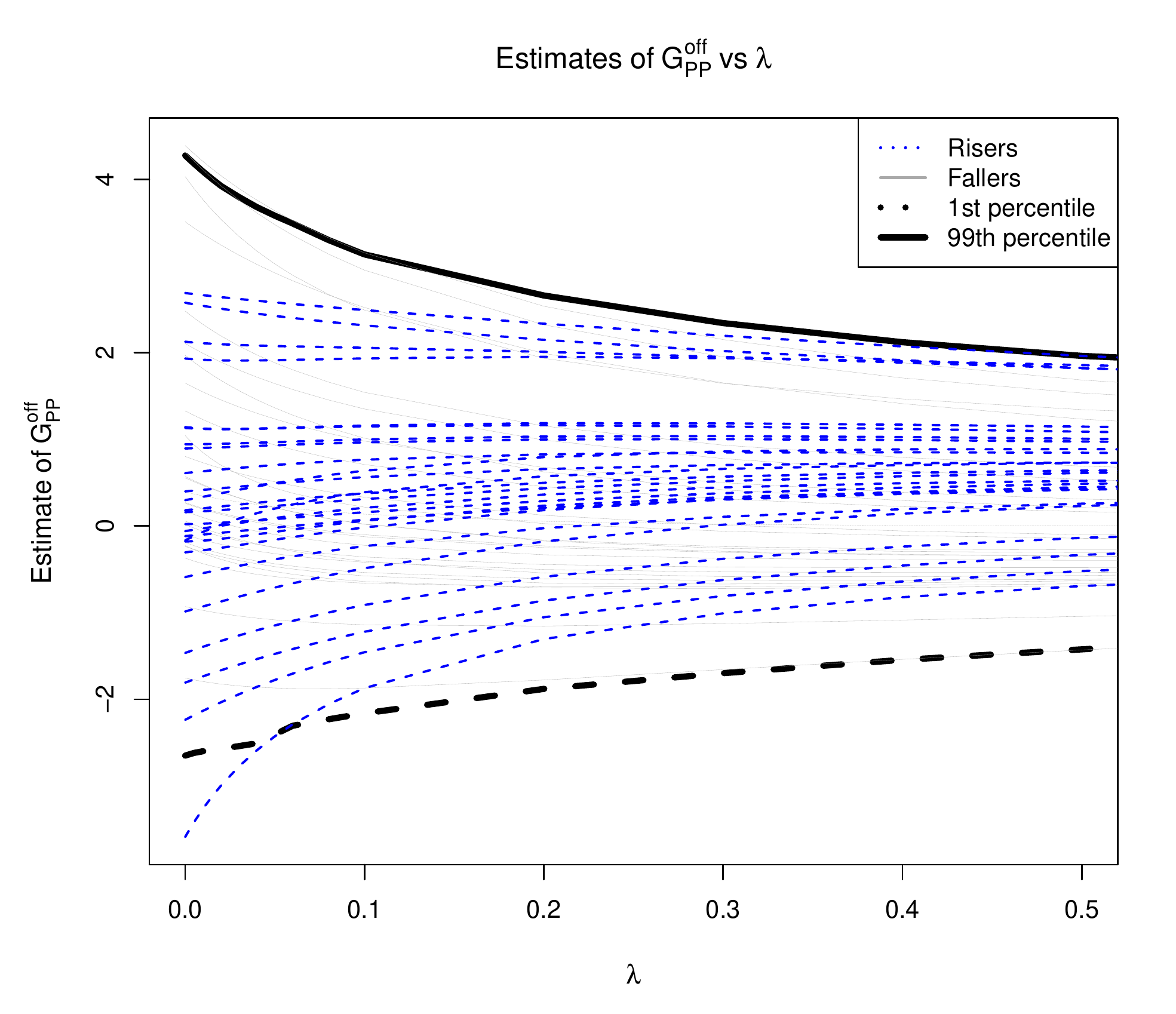}
            \caption[$G^\text{off}_{PP,60}$ vs $\lambda$ for players whose estimates change the most]{Estimates of $G^\text{off}_{PP,60}$ for different values of $\lambda$ for the 25 players whose coefficients were the most positively (resp. negatively) affected by the ridge regression, plotted as dashed (resp. solid) lines.}
            \label{tracecurves50}
            \end{figure}

        Datsyuk and Dubinsky are not the only players whose estimates exhibit this behavior.  We give the tracecurves of the 25 players whose coefficients were the most positively (resp. negatively) affected by the ridge regression as the dotted (resp. solid) lines in Figure \ref{tracecurves50}.  In many cases, there are drastic changes in a player's value relative the other players in the league.  A player may be worth 1 goal per 60 minutes more than another player according to their OLS estimates ($\lambda=0$) but worth 0.5 goals per 60 minutes \textit{less} according to their ridge estimates ($\lambda=0.5$). 

        \subsection{Year-to-year correlations}
\begin{figure}[h]
            \centering
            \includegraphics[width=.30\textwidth]{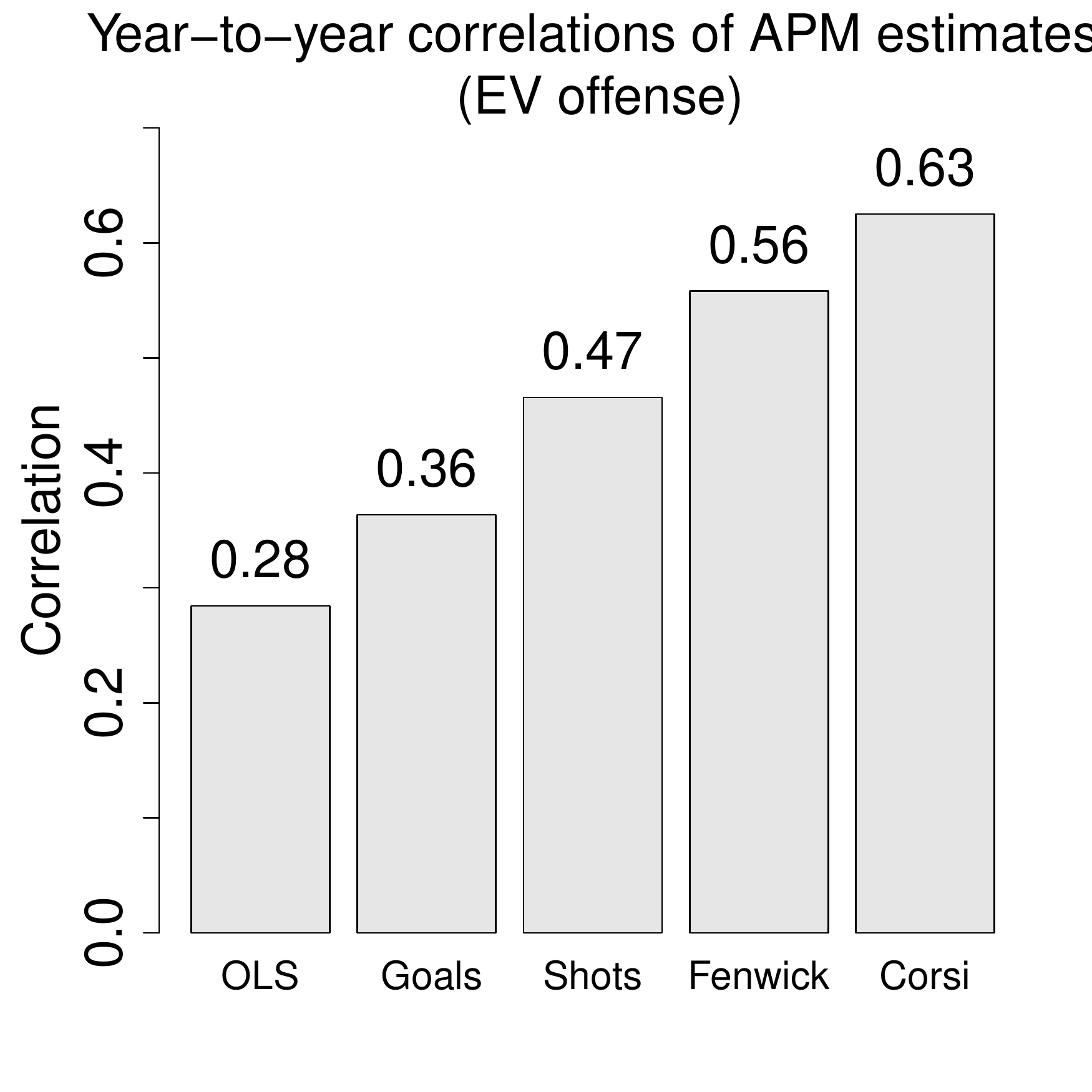}
            \includegraphics[width=.30\textwidth]{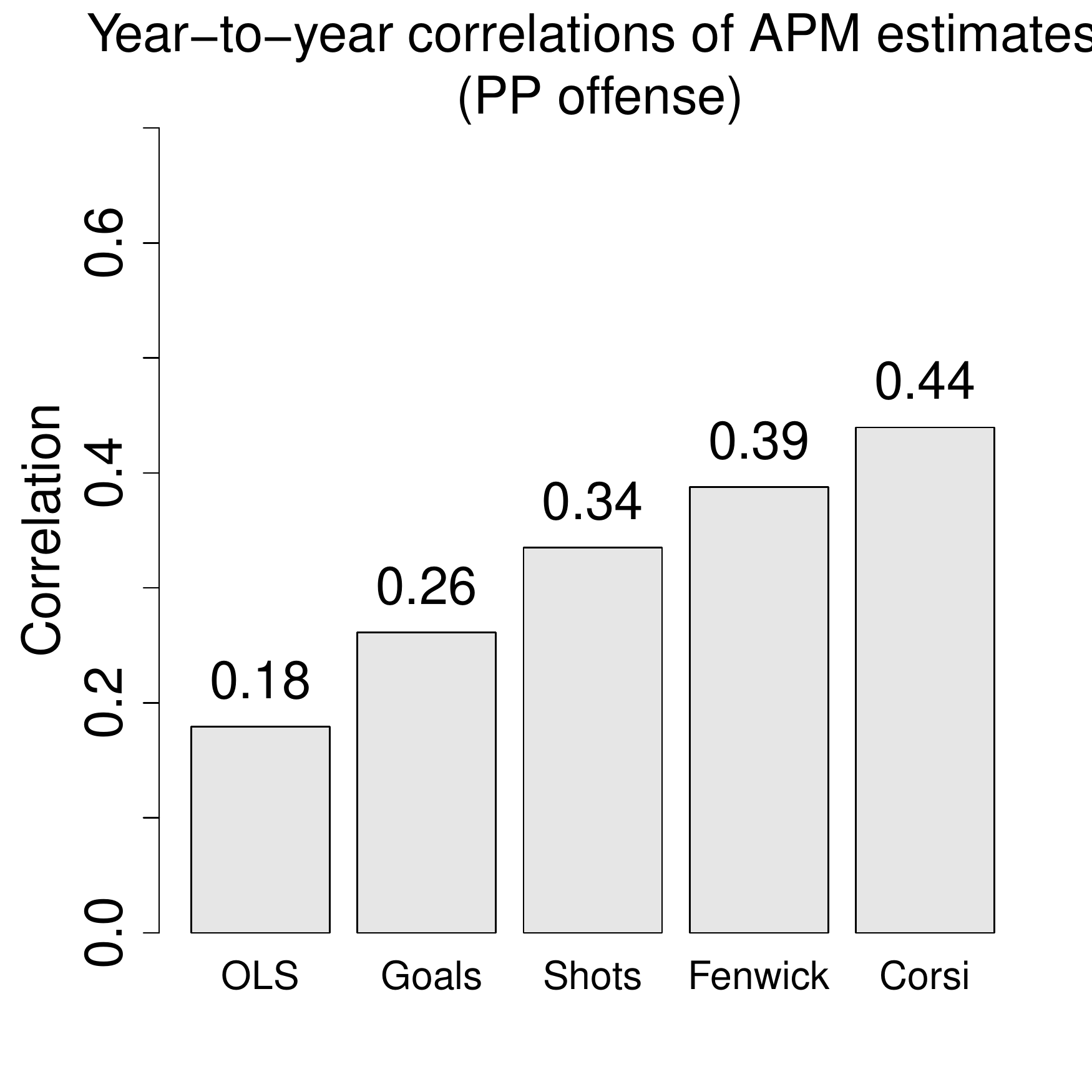}
            \includegraphics[width=.30\textwidth]{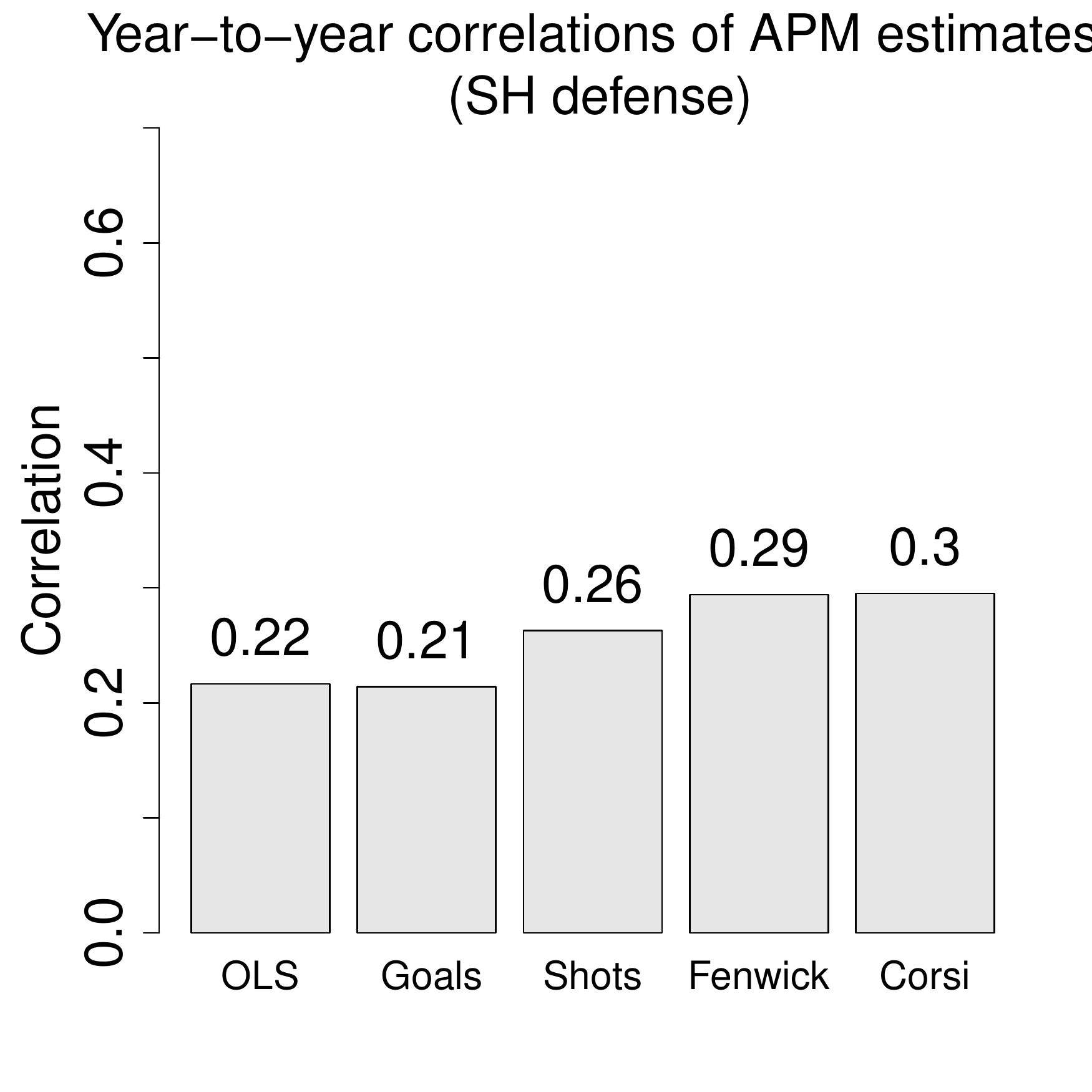}
            \caption[Year-to-year correlations of OLS and Ridge estimates]{A comparison of year-to-year correlations for OLS estimates (OLS) and our new ridge regression estimates (Goals, Shots, Fwick, Corsi). (Left) Even strength offense, minimum 500 minutes.  (Middle) Power play offense, minimum 150 minutes.  (Right) Short handed defense, minimum 150 minutes.  To compute the correlations, the per 60 minutes versions of these statistics were used.}
            \label{year-to-year-cor}
            \end{figure}

        We note that the ridge estimates tend to be more consistent from year to year than the OLS estimates.  In Figure \ref{year-to-year-cor} we give three examples of year-to-year correlations for three of the components of $APM$.  
        In the left figure, we see that our ridge estimates for offense at even strength using goals tend to be more consistent than the corresponding OLS estimates from \cite{apm} (which also used goals).  Also, the ridge estimates that use shots, Fenwick, and Corsi tend to be more consistent than the ridge estimates that use goals.  
        
        In the middle figure, we see that these trends are true for power play offense as well. 
        For short handed defense, the ridge estimates using goals are not more consistent than the OLS estimates, although the correlations for shots, Fenwick and Corsi are still higher.  We note that, in general, the even strength estimates tend to have higher year-to-year correlations than the power play and short handed estimates.  This trend is expected, since there is much less data for special teams situations than for even strength situations.  

\section{Results}\label{results}
    We now consider performance during the 2007-08, 2008-09, 2009-10, and 2010-11 seasons combined and determine the ``four-year'' Selke Trophy finalists (best defensive forwards), Norris Trophy finalists (best defensemen), and Hart Trophy finalists (most valuable players), according to $APM$.  Although the NHL typically announces three finalists for each trophy, we will give our top 5 finalists for each award and discuss other notable players.
   
\subsection{Four-year Selke Trophy finalists for Best Defensive Forward}
    Each season, the Selke Trophy is awarded to the forward that ``best excels at the defensive aspects of the game'' \cite{nhlcom}.  In practice, the award winner is typically a great defensive forward who contributes offensively as well.  In Table \ref{selke}, we give the top defensive forwards in the league during the 2007-08, 2008-09, 2009-10, and 2010-11 seasons according to $G^{\text{def}}_{}$.   
% latex table generated in R 2.13.1 by xtable 1.6-0 package
% Sun May 13 12:28:43 2012
\begin{table}[h!]
\begin{center}
\caption{The top 5 defensive forwards according to $G^{\text{def}}$.}
\label{selke}
{\footnotesize
\begin{tabular}{lllrrrrrrrr}
  \addlinespace[.3em] \toprule 
Player & Pos & Team & $G^{\text{def}}$ & $S^{\text{def}}$ & $F^{\text{def}}$ & $C^{\text{def}}$ & $G^{\text{def}}_{EV,60}$ & Err & $S^{\text{def}}_{EV,60}$ & Err \\ 
  \midrule 
Pavel Datsyuk & C & DET & 12 & 8 & 7 & 6 & 0.44 & 0.19 & 0.30 & 0.07 \\ 
  David Krejci & C & BOS & 11 & 3 & 2 & 0 & 0.52 & 0.20 & 0.18 & 0.07 \\ 
  Chris Higgins & LW & VAN & 10 & 0 & $-$1 & $-$2 & 0.33 & 0.21 & 0.03 & 0.07 \\ 
  Tomas Plekanec & C & MTL & 10 & $-$0 & $-$1 & $-$3 & 0.30 & 0.20 & 0.03 & 0.07 \\ 
  Mikko Koivu & C & MIN & 9 & 3 & 3 & 3 & 0.53 & 0.21 & 0.20 & 0.08 \\ 
   \bottomrule 
\end{tabular}
}
\end{center}
\end{table}
    Recall that $S^\text{off}$, $F^\text{off}$, $C^\text{off}$ have been rescaled by multiplying by the league average goals per shot, goals per Fenwick rating, and goals per Corsi rating, respectively.  Recall that these statistics are in the units of goals per season or expected goals per season based on shots, Fenwick, or Corsi.  

    Pavel Datsyuk seems to be the clear choice as the best defensive forward in the NHL according to $APM$.  He is the league leader in all 4 flavors of defensive contribution, and is also the best offensive player on the list.  The voters seem to agree: Datsyuk was awarded the Selke Trophy in 2007-08, 2008-09, and 2009-10, and he was a finalist in 2010-11.  Tomas Plekanec and Chris Higgins are on this list, but one might consider the next best candidates to be David Krejic and Mikko Koivu due to their superior ability to reduce the opposition's shots, Fenwick rating and Corsi rating.  Interestingly, multi-year finalist and 2010-11 winner Ryan Kesler is not on this list, although he did have very good numbers in 2010-11.  We note that 5 other players were tied with Koivu for 5th in $G^{\text{def}}$ with 9 to round out the top 10 in that category.  Daymond Langkow, who was 11th in $G^{\text{def}}$ with 8, missed the top 10 in $G^\text{def}_{}$, but was second in $S^\text{def}_{}$, and third in both $F^\text{def}_{}$ and $C^\text{def}_{}$.  In light of those rankings, Langkow could be considered one of the best defensive forwards in the game.  
    
\subsection{Four-year Norris Trophy finalists for Best Defensemen}
    The James Norris Memorial Trophy is given each year to the defensemen who ``demonstrates throughout the season the greatest all-round ability in the position'' \cite{nhlcom}.  In Table \ref{norris}, we give the top defensemen in the league during the 2007-08, 2008-09, 2009-10, and 2010-11 seasons according to $G$.
% latex table generated in R 2.13.1 by xtable 1.6-0 package
% Sun May 13 12:07:42 2012
\begin{table}[h!]
\begin{center}
\caption{The top 5 defensemen during the last 4 seasons, according to $G$.}
\label{norris}
{\footnotesize
\begin{tabular}{lllrrrrrrrrrr}
  \addlinespace[.3em] \toprule 
Player & Pos & Team & $G$ & $S$ & $F$ & $C$ & $G^{\text{off}}_{EV,60}$ & $S^{\text{off}}_{EV,60}$ & $G^{\text{off}}_{PP,60}$ & Err & $S^{\text{off}}_{PP,60}$ & Err \\ 
  \midrule 
Zdeno Chara & D & BOS & 19 & 9 & 9 & 10 & 0.10 & 0.21 & 0.43 & 0.33 & 0.58 & 0.07 \\ 
  Nicklas Lidstrom & D & DET & 19 & 1 & 3 & 5 & $-$0.06 & 0.07 & 1.37 & 0.26 & 0.71 & 0.05 \\ 
  Brian Campbell & D & CHI & 14 & 7 & 7 & 8 & 0.12 & 0.16 & 0.23 & 0.36 & 0.32 & 0.08 \\ 
  Andrei Markov & D & MTL & 13 & $-$3 & $-$1 & 1 & 0.20 & 0.13 & 1.75 & 0.37 & 0.59 & 0.08 \\ 
  Brian Rafalski & D & DET & 13 & 9 & 8 & 11 & $-$0.02 & 0.22 & 0.84 & 0.30 & 0.65 & 0.06 \\ 
   \bottomrule 
\end{tabular}
}
\end{center}
\end{table}
    It is not too surprising that Zdeno Chara and Nicklas Lidstrom were the best defenseman in the NHL during those seasons according to $G$.  Zdeno Chara's $APM$ results based on shots, Fenwick rating, and Corsi rating are better than those of Lidstrom, so one might choose to select him as the best defenseman.  Brian Campbell and Brian Rafalski are both strong across the board.    Interestingly, Andrei Markov does not rate very well in the $APM$ estimates based on shots, Fenwick rating, and Corsi rating.  One might prefer to include Chris Pronger, Dan Boyle, or Kris Letang instead of Markov on this list due to their ratings in $S, F$ and $C$.  Boyle, for example, led the league in $S$, $F$ and $C$.

\subsection{Four-year Hart Trophy finalists for Most Valuable Player}
    The Hart Memorial Trophy is given each year to the player ``judged to be the most valuable to his team'' \cite{nhlcom}.  Since $APM$ is not computed for goalies, we restrict our attention to only forwards and defensemen.  Typically, the Hart Trophy winner is a forward, in part because defensemen and goalies already have a trophy dedicated to the best player at those positions.  In Table \ref{hart}, we list the top 5 players in the league according to $G$.
% latex table generated in R 2.13.1 by xtable 1.6-0 package
% Sun May 13 12:13:43 2012
\begin{table}[h!]
\begin{center}
\caption{The top 5 players according to $G$.}
\label{hart}
{\footnotesize
\begin{tabular}{lllrrrrrrrrrr}
  \addlinespace[.3em] \toprule 
Player & Pos & Team & $G$ & $S$ & $F$ & $C$ & $G^{\text{off}}_{EV,60}$ & $S^{\text{off}}_{EV,60}$ & $G^{\text{off}}_{PP,60}$ & Err & $S^{\text{off}}_{PP,60}$ & Err \\ 
  \midrule 
Pavel Datsyuk & C & DET & 27 & 18 & 17 & 18 & 0.53 & 0.27 & 0.77 & 0.31 & 0.70 & 0.06 \\ 
  Jonathan Toews & C & CHI & 24 & 11 & 10 & 11 & 0.45 & 0.22 & 1.67 & 0.34 & 0.79 & 0.07 \\ 
  Alex Ovechkin & LW & WSH & 24 & 18 & 19 & 23 & 0.46 & 0.45 & 0.87 & 0.26 & 0.84 & 0.05 \\ 
  Daniel Sedin & LW & VAN & 23 & 16 & 16 & 17 & 0.47 & 0.44 & 1.11 & 0.26 & 0.73 & 0.06 \\ 
  Sidney Crosby & C & PIT & 22 & 11 & 12 & 12 & 0.83 & 0.42 & 0.98 & 0.29 & 0.58 & 0.06 \\ 
   \bottomrule 
\end{tabular}
}
\end{center}
\end{table}                  
    According to $G$, Pavel Datsyuk was the most valuable player in the league during the four seasons in question thanks to his excellent two-way play.  Datsyuk is also tied for first in $S$ and is third in $F$ and $C$.  
    
    Given the number of shots that Ovechkin throws at the net, it is not surprising that he is the leader in $S, F,$ and $C$, as well as the corresponding offensive components $S^\text{off}, F^\text{off},$ and $C^\text{off}$.  Ovechkin and Daniel Sedin have each won the Hart Trophy during the past four years, while Jonathan Toews has been a consistently excellent two-way player.  Toews has been a Selke finalist and a Conn Smythe trophy winner for the best player in the playoffs.  Unfortunately, Crosby missed significant time because of injury in two of the seasons that are used in this model.  Despite the injuries, Crosby still rates as the top offensive player in the league according to $G^\text{off}$, as we saw earlier in Table \ref{offense}. 
        
\section{Conclusions and Future Work}\label{conclusion}
    The use of ridge regression, and the addition of adjusted plus-minus models based on shots, Fenwick rating, and Corsi rating, are two valuable modifications of the earlier $APM$ models in hockey.  Other modifications could prove useful as well.  Different estimation techniques, such as that in \cite{thomas-ventura}, could be used.  Different outcome variables could also be used.  
    
    For example, one could also consider using weighted shots as the response variable in an $APM$ model.  By ``weighted shots'' we mean the following.  We could estimate the probability that a shot on goal will be a goal using distance, type of shot, and other details as explanatory variables.  Such shot quality models have been developed by Ken Krzywicki in \cite{ken1}, \cite{ken2}, and \cite{ken3} and Michael Schuckers in \cite{digr}.  Then, each shot can be weighted based on the probability that it will be a goal.  In a forthcoming article \cite{wshot}, the authors create a shot quality model similar to Krzywicki's logistic regression models, and use the resulting weighted shots as the outcome variable in a ridge regression model similar to the one described in this paper.  The results of this model are estimates of $W$, an adjusted plus-minus rating based on weighted shots.  
    
    Also, recall that Fenwick rating and Corsi rating are combinations of shots, missed shots, and blocked shots, and are a good indication of possession advantage and team performance in general.  One could build on the idea of using those statistics and consider other statistics like hits, faceoffs, and zone starts as well.  In \cite{spm}, the author estimates the combinations of these statistics are the best predictors of goal scoring at the team level.  The results of the model can be interpreted as ``expected goals''.  These expected goals are then used as the outcome in a ridge regression similar to the model described in this paper.  The results are estimates of $E$, an adjusted plus-minus rating based on expected goals.  Another approach that uses several different statistics can be found in \cite{lockschuckers}.
    
    We hope that the ideas presented in this paper will be useful to fans, analysts, coaches and teams as they analyze the performance of NHL players,  
    and will inspire future work in performance analysis in hockey.

\section*{Acknowledgements}
I would like to thank William Pulleyblank for many useful conversations about this work and for his comments and suggestions after reading a draft of the paper.  I would also like to thank the referees for many comments and suggestions that improved the quality of this paper.

%\newlength{\bibhang}
\setlength{\bibhang}{.5in} % indents the second lines of the bib
\bibliographystyle{DeGruyter}
\bibliography{generalbib}

\section{Appendix}\label{app}

\subsection{Ordinary Least Squares}
        To find the ``best fit'' solution of \eqref{matrixform} using ordinary least squares (OLS) regression, 
        one finds the $\beta_j$s, $\delta_j$s, and $\zeta$s that minimize the sum of squared error
        \begin{align}\label{olsSSE}
        Q = \sum_{i=1}^N (y_i - \hat{y_i})^2, 
        \end{align}
        where $\hat{y_i}$ is the predicted outcome for observation $i$ and is given by
        \begin{align}\label{yhat}
        \hat{y_i} = \beta_0 &+ \beta_1 X_{1,i} + \cdots + \beta_J X_{J,i} %\notag \\
                                                 + \delta_1 D_{1,i} + \cdots + \delta_J D_{J,i} 
                                                  + \zeta_{off} Z_{off,i} + \zeta_{def} Z_{def,i}.
        \end{align}
        In matrix notation, the sum of squared error $Q$ in \eqref{olsSSE} can be written 
            \begin{equation}\label{olsmatrix}
            Q = (y - X\beta)^T(y-X\beta),
            \end{equation} 
        where $(y - X\beta)^T$ denotes the transpose of $y - X\beta$.  Equivalently, finding the least squares estimates of $\beta$ amounts to finding the $\hat{\beta}$ that solves the system
            \begin{align}\label{normalequation}
                X^T X \hat{\beta} = X^T y,
            \end{align}
        which is obtained by multiplying both sides of \eqref{matrixform} by $X^T$ on the left.  When there is only one predictor variable, finding $\hat\beta$ can be thought of as finding the line that best fits the data.  With two predictor variables, one finds the plane that best fits the data.  With more than two variables, the case we have in this paper, one finds the best fit hyperplane.  
        
        If the kernel (or nullspace) of $X$ is 0, which is typically true when $N>> J$, then $X^T X$ is invertible, and we can solve for $\hat{\beta}$ by multiplying both sides of \eqref{normalequation} by $(X^T X)^{-1}$ on the left, giving
            \begin{align}
                \hat{\beta} &= (X^T X)^{-1} X^T y. \notag
            \end{align}
        Further details about OLS from a linear algebraic point of view can be found in most standard undergraduate linear algebra textbooks (for example, \cite{strang}, \cite{bretscher}, or \cite{lay}) or a multiple linear regression textbook (for example, \cite{applied-linear-regression-models}).
            
        Ordinary least squares was the approach taken in \cite{apm} and \cite{apm2} and several of the basketball articles.  Unfortunately, collinearity in $X$ results in high standard errors for $\hat{\beta}$.  A linear algebraist might prefer the viewpoint that if two teammates play together often, then two columns of $X$ are nearly the same, the columns of $X$ are nearly linearly dependent, and the corresponding columns (and rows) $X^T X$ are nearly linearly dependent, which means that $X^T X$ is nearly singular and has a high condition number.  A high condition number means that solutions to \eqref{normalequation} are sensitive to small changes in the data, an undesirable property.  It also leads to large standard errors in the estimates of $\beta.$ 

    \subsection{Ridge Regression}
        In ridge regression, instead of finding the $\beta$ that minimizes \eqref{olsmatrix}, one standardizes the columns of $X$ and finds the $\beta$ that minimizes
            \begin{equation}\label{ridgeSSE}
            Q =  (y - X\beta)^T(y-X\beta) + \lambda \beta^T \beta
            \end{equation}
        where $\lambda$ is a ridge parameter that needs to be chosen.  Note that \eqref{ridgeSSE} is similar to \eqref{olsmatrix} but with the penalty term $\lambda \beta^T \beta$ included.  Equivalently, instead of solving \eqref{normalequation} for $\hat\beta$, one solves the equation
            \begin{align}\label{ridgeequation}
                (X^T X + \lambda I)\hat{\beta} = X^T y,
            \end{align}
         for $\hat\beta$, where $I$ denotes the identity matrix.  Note that \eqref{ridgeequation} is similar to \eqref{normalequation} but with the penalty term $\lambda I$ included.  
         
         To solve \eqref{ridgeequation}, one multiples both sides of the equation by $(X^T X + \lambda I)^{-1}$ on the left, which gives
                     \begin{align}\label{ridgeequation}
                         \hat{\beta} &= (X^T X + \lambda I)^{-1} X^T y. 
                     \end{align}
                  These estimates $\hat{\beta}$ are the estimates that we use in the next section to evaluate players.  The interpretation of $\hat\beta$ is the same for ridge regression as it was with OLS regression.  In our case, coefficients $\hat\beta$ are estimates of the offensive and defensive contributions of players in terms of goals per 60 minutes, independent of the strength of their teammates and opponents, and independent of the zone in which their shifts begin.
         
         The effect of the penalty term is to penalize large values for the coefficients $\beta$.   Ridge regression can be thought of like OLS regression, which finds the ``best fit'' hyperplane, but with constraints on the coefficients $\beta$ that prevent them from being poorly behaved.  Note that for the choice $\lambda=0$, \eqref{ridgeSSE} becomes \eqref{olsmatrix}, and \eqref{ridgeequation} becomes \eqref{normalequation}, so $\lambda=0$ in ridge regression corresponds to the ordinary least squares estimates, where the coefficients are unconstrained and may have high error bounds.  As $\lambda$ increases, the coefficients tend to stabilize and move toward zero.  
         
         We remark that including the penalty term $\lambda I$ in \eqref{ridgeequation} can seem somewhat ad hoc or arbitrary, but fortunately there is a nice Bayesian justification for this approach.  The ridge regression model \eqref{ridgeequation} is equivalent to a Bayesian regression model in which the coefficients $\beta$ are given a normal prior distribution with mean 0 and a variance that depends on $\lambda$.  Changing $\lambda$ corresponds to changing how influential the mean 0 prior will be on the value of the estimates.  From a linear algebra perspective, the term $\lambda I$ is effectively padding the diagonal of $X^T X$, which lowers its condition number, and makes the solutions $\hat{\beta}$ less volatile. 

        \subsection{Choosing $\lambda$}
        Often, the ridge parameter $\lambda$ is chosen using cross-validation.  With large data, specifically when $n$, the number of rows, is large, computing $\lambda$ in this way can be computationally expensive, as it requires one to compute $n$ leave-one-out estimates.  Another alternative is generalized cross-validation (GCV), which is also computationally expensive.  To see why, consider the hat matrix 
            \begin{equation}\label{hatmatrix}
                    H = X(X^TX + \lambda I)^{-1} X^T.  
            \end{equation}    
        Finding $H$, or the trace of $H$, is a required step for GCV.  If $X$ has $n$ rows, then $H$ is an $n$-by-$n$ matrix.  For our even strength model, for example, we have well over 1,000,000 rows, meaning $H$ is a 1,000,000 by 1,000,000 matrix.  
        
        In our work, we use an estimate of the trace of $H$ to get a randomized version of GCV simliar to that in \cite{girard}.  This method uses the following lemma given in \cite{hutchinson}:
            \newtheorem*{thmnonum}{Lemma}
            \begin{thmnonum}[\cite{hutchinson}]
                Let $B$ be an $n\times n$ symmetric matrix and let $u = (u_1, \ldots, u_n)^T$ be a vector of $n$ independent samples from a random variable $U$ with mean 0 and variance $\sigma^2$.  Then,
                    \begin{eqnarray}\label{thmequ}
                    E(\epsilon^T B \epsilon) = \sigma^2\textup{tr}(B).
                    \end{eqnarray}  
            \end{thmnonum}
        Note that $E(\cdot)$ denotes expectation and $\text{tr}(\cdot)$ denotes the trace of a matrix. 
        The hat matrix $H$ is symmetric, so the lemma applies.  The lemma is useful because $\frac{1}{\sigma^2} \epsilon^T H \epsilon$ is easier to compute than tr$(H)$, and $\frac{1}{\sigma^2} \epsilon^T H \epsilon$ is an unbiased estimate for tr$(H)$ according to the lemma.  Also, the estimate is very accurate (see, for example \cite{girard} or \cite{hutchinson}). 
        
        Note that using \eqref{thmequ} and \eqref{hatmatrix} we can write 
            \begin{equation}\label{ehe}
                \textup{tr}(H) \approx \frac{1}{\sigma^2}\epsilon^T H \epsilon =\frac{1}{\sigma^2} \epsilon^T [X(X^TX + \lambda I)^{-1} X^T] \epsilon 
            \end{equation}
        and since matrix multiplication is associative, we can group the terms in $$\epsilon^T [X(X^TX + \lambda I)^{-1} X^T] \epsilon $$ in any order.  We can write \eqref{ehe} as 
            \begin{equation}
                 \textup{tr}(H) \approx \frac{1}{\sigma^2}(\epsilon^T X)(X^TX + \lambda I)^{-1}(X^T \epsilon).
                 \label{traceHest}
            \end{equation}
        Note that if $X$ is an $n\times p$ matrix and $\epsilon$ is an $n\times 1$ matrix, then \begin{align*}
            \epsilon^T X     &\quad \text{is a  } 1 \times p \text{ matrix}, \\
            X^TX + \lambda I &\quad \text{is a  } p \times p \text{ matrix, and }  \\
            X^T\epsilon      &\quad \text{is a  } p \times 1 \text{ matrix}, 
        \end{align*}
        so our biggest matrix is $p \times p$.  Since typically $p<<n$ when $n$ is very large, it is much easier to work with a $p\times p$ matrix than an $n\times n$ matrix.  In our case, for example, $n$ is on the order of 1,000,000, while $p$ is on the order of only 1,000.  We used the estimate for the trace in \eqref{traceHest} to obtain a randomized GCV choice for $\lambda$ as in \cite{girard}.
            
        In some cases, we preferred to increase the value of $\lambda$ obtained by this method.  This change can be justified in several ways.  
        %First, increasing $\lambda$ leads to more conservative estimates, by which we mean estimates that are closer to zero.  
        For example, in some cases, inspection of the trace curves (that is, the curves like those in Figure \ref{tracecurves}) revealed that the estimates did not yet appear to be stabilized at those values of $\lambda$, and this observation can be used to justify increasing $\lambda.$
                We also  considered the Hoerl-Kannard-Baldwin estimate \cite{hoerl-kannard-baldwin} of $\lambda$.  The Hoerl-Kennard-Baldwin estimate is given by 
                    \begin{equation}
                        \lambda_{HKB} = \frac{p \, \, MSE}{\hat{\beta}^T\hat{\beta}},
                    \end{equation}
                where MSE denotes mean-squared error.  Finally, we considered variance inflation factors (VIF), which quantify the level of collinearity present in the data, when choosing $\lambda.$
%        Let $MSE$ denote mean-squared error, let Var $\beta$ be the variance-covariance matrix of $\beta$, and let diag(Var $\beta$) denote the diagonal entries of that matrix.  Then the variance inflation factors are defined as
%            \begin{equation}
%            \frac{\text{diag}(\text{Var } \hat{\beta})}{MSE}
%            \end{equation}
%        It was shown in \cite{hoerl-kennard-ridge} that
%        
%            \begin{equation}
%                \text{Var } \hat{\beta} = \sigma^2 (X^T X + \lambda I)^{-1}X^T X (X^T X + \lambda I)^{-1}
%            \end{equation}
        As stated in \cite{marquardt}, the VIF can be expressed as the diagonal elements of 
                    \begin{equation}
                        (X^T X + \lambda I)^{-1}X^T X (X^T X + \lambda I)^{-1}.
                    \end{equation}
        Typically, values in the single digits are preferred.  Often the VIF were high for the values of $\lambda$ that we got using GCV.  We chose $\lambda$ at least high enough so that the VIF were below 10.
        %so that slightly, the VIF were drastically reduced and single digit values were obtained.  We increase
        
        These four pieces of information were considered when choosing $\lambda$ for each of our 8 models that used 4 seasons of data from the 2007-08 through 2010-2011 seasons.  We also used this information with models that only used a single season's worth of data, giving 8 more values of $\lambda$ for each season.  In each case, we chose the highest value of $\lambda$ suggested by these four methods.  These values of $\lambda$ were used in \eqref{ridgeequation} to obtain estimates of the coefficients in each of our models.  The vertical line at $\lambda = 0.5$ in Figure \ref{tracecurves} indicates the value of $\lambda$ that we chose for that model.  Note that the estimates seem to have stabilized for the most part by the time $\lambda$ reaches 0.5.  
        %The values of $\lambda$ for the other models ranged between 0.1 and 0.75. In general, the ridge parameter was greater for the models with less data (single season models, models with goals, models for special teams).   

    \subsection{Supplemental figures}
            \newpage
            \vfill
            \bigskip
            \vfill
\end{document}